\documentclass[conference]{IEEEtran}
\usepackage[utf8]{inputenc}
\IEEEoverridecommandlockouts
\usepackage{cite}
\usepackage{amsmath,amssymb,amsfonts}
\usepackage{algorithmic}
\usepackage{graphicx}
\usepackage{textcomp}
\usepackage{stfloats}
\def\BibTeX{{\rm B\kern-.05em{\sc i\kern-.025em b}\kern-.08em
    T\kern-.1667em\lower.7ex\hbox{E}\kern-.125emX}}
    
\begin{document}

\title{A large-scale evaluation framework for EEG deep learning architectures\\
\thanks{This work was supported by DFG grant EXC1086 BrainLinks-BrainTools, Baden-Württemberg Stiftung grant BMI-Bot, Graduate School of Robotics in Freiburg, Germany and the State Graduate Funding Program of Baden-Württemberg, Germany.}
}

\author{

    \IEEEauthorblockN{Felix A. Heilmeyer}
    \IEEEauthorblockA{Translational Neurotechnology Lab\\
    University Medical Center Freiburg\\
    Freiburg, Germany\\
    science@fehe.eu}
    
    \and
    
    \IEEEauthorblockN{Robin T. Schirrmeister}
    \IEEEauthorblockA{Translational Neurotechnology Lab\\
    University Medical Center Freiburg\\
    Freiburg, Germany\\
    robin.schirrmeister@uniklinik-freiburg.de}
    
    \and
    
    \IEEEauthorblockN{Lukas D. J. Fiederer}
    \IEEEauthorblockA{Translational Neurotechnology Lab\\
    University Medical Center Freiburg\\
    Freiburg, Germany\\
    lukas.fiederer@uniklinik-freiburg.de}
    
    \and

    \IEEEauthorblockN{Martin Völker}
    \IEEEauthorblockA{Translational Neurotechnology Lab\\
    University Medical Center Freiburg\\
    Freiburg, Germany\\
    martin.voelker@uniklinik-freiburg.de}
    
    \and
    
    \IEEEauthorblockN{Joos Behncke}
    \IEEEauthorblockA{Translational Neurotechnology Lab\\
    University Medical Center Freiburg\\
    Freiburg, Germany\\
    joos.behncke@uniklinik-freiburg.de}
    
    \and
    
    \IEEEauthorblockN{Tonio Ball}
    \IEEEauthorblockA{Translational Neurotechnology Lab\\
    University Medical Center Freiburg\\
    Freiburg, Germany\\
    tonio.ball@uniklinik-freiburg.de}
}

\maketitle

\begin{abstract}
EEG is the most common signal source for noninvasive BCI applications. For such applications, the EEG signal needs to be decoded and translated into appropriate actions. A recently emerging EEG decoding approach is deep learning with Convolutional or Recurrent Neural Networks (CNNs, RNNs) with many different architectures already published. Here we present a novel framework for the large-scale evaluation of different deep-learning architectures on different EEG datasets. This framework comprises (i) a collection of EEG datasets currently including 100 examples (recording sessions) from six different classification problems, (ii) a collection of different EEG decoding algorithms, and (iii) a wrapper linking the decoders to the data as well as handling structured documentation of all settings and (hyper-) parameters and statistics, designed to ensure transparency and reproducibility. As an applications example we used our framework by comparing three publicly available CNN architectures: the Braindecode Deep4 ConvNet, Braindecode Shallow ConvNet, and two versions of EEGNet. We also show how our framework can be used to study similarities and differences in the performance of different decoding methods across tasks. We argue that the deep learning EEG framework as described here could help to tap the full potential of deep learning for BCI applications.

\end{abstract}

\begin{IEEEkeywords}
EEG, BCI, Deep Learning, Convolutional Neural Networks, Braindecode, EEGNet, FBCSP, Performance Comparison
\end{IEEEkeywords}

\section{Introduction}
EEG is the most common signal source for noninvasive BCI applications. For such applications to work reliably, the EEG signal needs to be decoded with high accuracy and translated into appropriate actions. To this purpose a large and growing variety of decoding methods is being used. A recently emerging EEG decoding approach is deep learning with Convolutional or Recurrent Neural Networks (CNNs, RNNs). Deep learning has already revolutionized other areas and excels at decoding information from raw data, e.g., in image recognition \cite{Krizhevsky:2012wl} and natural language processing \cite {Kim:2014vt}. Thus, currently many BCI researchers are starting to investigate the potential usefulness of deep learning techniques using a wide range of different network architectures and applying them to a wide range of EEG datasets \cite{Bashivan:2015wm,Tang:2017kq,Lee:bx,Schirrmeister:2017bv,burget_acting_2017}.

Faced with this large variety in architectures and applications, choosing a network architecture for new BCI tasks is not trivial for various reasons. Although most of the published studies evaluated the performance of their deep learning architectures against some comparison algorithm, in most of the cases these comparisons were either against traditional, non-deep-learning decoding methods, or involved different versions of the newly introduced network architecture. Studies evaluating against other, already published CNN- or RNN-based analyses are less common. Moreover, the number of different datasets used in these evaluations is often small and may not reflect the wide range of EEG decoding problems of different difficulty \cite{Schirrmeister:2017bv,Lawhern:2018fb,Lawhern:2016ta}.

Therefore, a framework for the systematic evaluation of deep learning for EEG which addresses these challenges is desirable. As a first step in this direction, we developed a framework to compare deep learning architectures on a large set of EEG examples ($>$100 decoding problems) in a comprehensively documented and reproducible manner. Applying this framework, here we report on three publicly available CNN architectures.

In this paper we describe the framework and provide our rationale for design choices regarding the challenges describes above. Secondly, we present the results of our comparison and make a recommendation on which of the network(s) included in the comparison to use for best performance. At last we discuss how the present framework could be further extended and improved.

\section{Methods}
Our framework is build upon three components: (i) a collection of EEG data, (ii) the decoding methods embedded in the Braindecode toolbox, and (iii) a wrapper that enables running large-scale decoding experiments in an easily reproducible fashion. This section contains a detailed description of these three components.

\subsection{EEG Data}
The performance evaluation was done on a range of datasets representing a spectrum of common BCI tasks including motor tasks, speech imagery, and error processing. We choose to include different tasks to ensure that success or failure of the compared methods is not limited to a specific decoding domain.
The difficulty of the included decoding problems ranges from data which can be decoded by all methods with high accuracy to data which is almost impossible to decode with currently available methods. One might argue that data too difficult to decode for all employed methods does not contribute to the comparison. We still included this data for the following reason: Assume we only included easy to decode data, on which all methods already achieve high decoding accuracies. In this scenario, if repeating the evaluation in the future with a potentially better performing method, that method could not show its full potential. In contrast, by including difficult data, on which current methods only achieve medium accuracies, we leave room for future methods to show their superiority. This allows us to subsequently expand this comparison to new emerging decoding methods. 

All included datasets were acquired at the Translational Neurotechnology Lab, University of Freiburg, and recorded with an EEG cap with 128 gel-filled electrodes. During recording subjects were presented with different stimuli or had to perform specific tasks. The following paragraphs contain a short description of the six paradigms included in this study. For a detailed description of the data acquisition and experimental setups please refer to the respective cited original publications. Together, the datasets amounted to 100 different decoding task examples. Table \ref{tabmeta} gives an overview of the included datasets.

\begin{table*}[ht]
\caption{Dataset summary}
\begin{center}
\begin{tabular}{|c|c|c|c|c|c|}
\hline
\textbf{Name (Acronym)}&\textbf{\# Classes}&\textbf{Task Type}&\textbf{\# Subjects}&\textbf{Trials per Subject}&\textbf{Class balance} \\
\hline
High-Gamma Dataset (Motor) & 4 & Motor task & 20 & 1000 & balanced\\
KUKA Pouring Observation (KPO) & 2 & Error observation & 5 & 720-800 & balanced \\ 
Robot-Grasping Observation (RGO) & 2 & Error observation & 12 & 720-800 & balanced \\ 
Error-Related Negativity (ERN) & 2 & Eriksen flanker task & 31 & 1000 & 1/2 up to 1/15\\ 
Semantic Categories & 3 & Speech imagery & 16 & 750 & balanced\\ 
Real vs. Pseudo Words & 2 & Speech imagery & 16 & 1000 & 3/1\\
\hline
Total &  &  & 100 & & \\
\hline
\end{tabular}
\label{tabmeta}
\end{center}
\end{table*}

Our first dataset was initially published in \cite{Schirrmeister:2017bv} and named "High-Gamma Dataset" (motor) as the recording setup was optimized to capture movement related frequencies in the high gamma range. For recording, subjects were instructed via visual stimuli to hold still, tap the fingers of either left hand, right hand, or flex the toes of both feet. The decoding task evaluated in this study is to classify which of those four instructions was executed at each trial. Recording data from stimulus onset to 4\,s after onset was used for decoding.

Our second dataset dubbed "Error-Related Negativity" (ern) used a variant of the Eriksen flanker task\cite{Eriksen:1979ex}. This involves reacting as fast as possible to a visual stimulus by pressing a button with either the left or right index finger. When subjects reacted to the stimulus with pressing the correct button, they were rewarded with points; When they pressed the wrong button, they were penalized by losing points. The according decoding problem is to classify for each trial if the subject was successful or failed, that is pressing the correct or the wrong button. The data used for decoding in this study was the EEG recorded from 0.5 s before to 1.5 s after a button was pressed. A detailed description can be found in \cite{volker2018dynamics}.

In the KUKA Pouring Observation paradigm (kpo) subjects were watching a video of a robotic arm pouring liquid from a bottle into a glass. In each trial the robot either succeeded or spilled the liquid. On this data the task for the decoder is to classify whether the subject was watching a video of a successful or unsuccessful attempt.

The Robot-Grasping Observation paradigm (rgo) consisted of subjects watching a video of a robot approaching, grabbing and then lifting a ball from the floor. Similar to the previous experiment, in each task the robot was either failing or succeeding; And the decoding task is to classify whether the subject was watching failure or success.

In both observation paradigms the error occurred in an interval from 2.5\,s to 5\,s after stimulus start. Consequently, data recorded in this interval was used for decoding. Publications featuring detailed descriptions of these two datasets are \cite{Behncke:2017ug, Welke:2017tg}.

The last dataset used in this study concerned semantic processing. It was first published in \cite{Rau:2015uk}. The recording setup matches the setup used for recording the previous datasets. The subjects were presented with a word on a computer screen for 500\,ms and instructed to repeat the word silently for three seconds following the stimulus. Data recorded in these three seconds was later utilized for decoding. The presented words were 84 concrete nouns of three semantic categories: food, animals, and tools. Additionally, an equal number of pseudowords was included. These pseudowords were constructed to look and feel like real words but carry no meaning \cite{Blanken:1999wk}. We created two separate decoding tasks from this dataset, one with three and one with two classes, respectively, by (i) labeling trials with one of the three semantic categories (omitting the pseudowords) and by (ii) distinguishing real vs pseudowords (semantic and pseudovsreal, respectively).\bigskip

The following paragraphs outline how the data was preprocessed before decoding and why this specific preprocessing approach was chosen.

The preprocessing was the same for all methods and all datasets. Most likely, decoding performance would have profited from individualized preprocessing. Nevertheless, we choose to use a uniform preprocessing to ensure that observed performance differences were the effect of the decoding method alone.
First, the data was downsampled to 250\,Hz and bandpass filtered with a 0.5–120 Hz filter. This preserves the main neurophysiologically important frequency bands usually considered in EEG recordings but also reduces the amount of data to allow reasonable training duration, even with deep CNNs.

At this point the last 20\% of every subject's recording were split and put aside for final testing. The cleaning procedures described in the following were only applied to the training data as it is customary in developing many machine learning applications. Thereby, the testing results reflect actual decoding performance when the trained classifier is applied to new data without cleaning. We restricted cleaning to the recording artifacts employing the following algorithm: First, following our procedure as described in \cite{Schirrmeister:2017bv}, all channels in which more than 20\% of the samples were over 800 µV were marked as broken and removed. Then trials were cut and any trial which still contained samples over 800 µV was removed. This simple cleaning mechanism thus removed large-amplitude artifacts that would likely disturb the training process.

\begin{table*}[ht]
\centering
\caption{Classifier Architectures}
\begin{tabular}{|l|l|l|l|l|}
\hline
\textbf{}        & \textbf{Deep}                  & \textbf{Shallow}         & \textbf{EEGNet v1}      & \textbf{EEGNet v2}                \\ \hline
\textbf{Layer 1} & 25xConv2D (10, 1), Stride 1,1  & 40xConv2D (25, 1), St1,1 & 16xConv1D (C, 1), St1,1 & 8xConv2D (1,64), St1,1            \\
\textbf{}        & 25xConv2D (1, C), Stride 1,1   & 40xConv2D (1, C), St1,1  & BatchNorm               & BatchNorm                         \\
\textbf{}        & BatchNorm                      & BatchNorm                & Activation (ELU)        & 16xDepthwiseConv2D (C,1), St1,1   \\
\textbf{}        & Activation (ELU)               & Activation (Square)      & Transpose to 16,T       & BatchNorm                         \\
\textbf{}        & MaxPool (3, 1), Stride 3,1     & MeanPool (75, 1), St15,1 & Dropout (0.25)          & Activation (ELU)                  \\
\textbf{}        &                                & Activation (Log)         &                         & AveragePool2D (1,4), Stride 1,1   \\
\textbf{}        &                                & Dropout (0.5)            &                         & Dropout (0.25)                    \\ \hline
\textbf{Layer 2} & Dropout (0.5)                  & Dense                    & 4xConv2D (2,32), St1,1  & 16xSeparableConv2D (1,6), Str1, 1 \\
\textbf{}        & 50xConv2D (10, 1), Stride 1,1  & Softmax Classification   & BatchNorm               & BatchNorm                         \\
\textbf{}        & BatchNorm                      &                          & Activation (ELU)        & Activation (ELU)                  \\
\textbf{}        & Activation (ELU)               &                          & MaxPool (2,4), St1,1    & AveragePool2D (1,8), St1,1        \\
\textbf{}        & MaxPool (3, 1), Stride 3,1     &                          & Dropout (0.25)          & Dropout (0.25)                    \\ \hline
\textbf{Layer 3} & Dropout (0.5)                  &                          & 4xConv2D (8,4), St1,1   & Dense                             \\
\textbf{}        & 100xConv2D (10, 1), Stride 1,1 &                          & BatchNorm               & Softmax Classification            \\
\textbf{}        & BatchNorm                      &                          & Activation (ELU)        &                                   \\
\textbf{}        & Activation (ELU)               &                          & MaxPool (2,4), St1,1    &                                   \\
\textbf{}        & MaxPool (3, 1), Stride 3,1     &                          & Dropout (0.25)          &                                   \\ \hline
\textbf{Layer 4} & Dropout (0.5)                  &                          & Dense                   &                                   \\
\textbf{}        & 200xConv2D (10, 1), Stride 1,1 &                          & Softmax Classification  &                                   \\
\textbf{}        & BatchNorm                      &                          &                         &                                   \\
\textbf{}        & Activation (ELU)               &                          &                         &                                   \\
\textbf{}        & MaxPool (3, 1), Stride 3,1     &                          &                         &                                   \\ \hline
\textbf{Layer 5} & Dense                          &                          &                         &                                   \\
                 & Softmax Classification         &                          &                         &                                   \\ \hline
\end{tabular}
\label{tabarchitectures}
\end{table*}

\subsection{Decoding methods}
In this section we will shortly introduce the decoding methods compared in this study. For an overview of all included architectures refer to Table \ref{tabarchitectures}. To avoid errors we used the implementations of the original authors for all decoding methods. For implementation details refer to the respective publications.

Firstly, we used two CNN architectures that are part of the Braindecode open source toolbox for EEG decoding released last year by our lab and published in \cite{Schirrmeister:2017bv}: The Braindecode Deep4 ConvNet and Braindecode Shallow ConvNet, hereafter referred to as Deep4 Network and Shallow Network. The Deep4 Network features four convolution-max-pooling blocks, using batch normalization and dropout, followed by a dense softmax classification layer. The first block is split, first performing a temporal convolution then a spatial convolution over all channels followed by the max pooling. The other three blocks are standard convolution-max-pooling blocks. All layers use exponential linear units (ELUs) as nonlinearities.

The Shallow Network architecture also features a temporal then a spatial convolution layer, followed by a squaring nonlinearity, a mean-pooling layer and a dense classification layer. This architecture has many similarities to the FBCSP method. For further details refer to \cite{Schirrmeister:2017bv}.

The third decoding method in this comparison was a CNN called EEGNet\cite{Lawhern:2016ta} designed for compactness (few trainable parameters). It consists of three convolutional layers followed by a softmax regression layer. The first convolutional layer only convolutes spatially; Following layers also convolute temporal and include max pooling. Additionally, every layer uses batch normalization and dropout.

Shortly before the submission of this paper the authors of EEGNet published an updated version of EEGNet with a fundamentally changed architecture\cite{Lawhern:2018fb}, which we will refer to as EEGNetv2 in this paper. The new EEGNetv2 architecture uses first temporal then spatial convolutions in its first layer. Hence the initial layers are now more similar to the Braindecode networks compared to the original EEGNet. In contrast to the Braindecode networks EEGNetv2 uses a depthwise convolution with a depth multiplier of two, therefore learning two separate spatial filters for each location. Additionally, the second layer of EEGNetv2 uses separable instead of standard convolutions. Lastly, EEGNetv2 uses average pooling, whereas the original EEGNet uses max pooling. For an overview of the architecture differences refer to Table \ref{tabarchitectures}.

\subsection{Comparison wrapper, statistics and correlation analysis}
The last component of our framework is a wrapper that enables large scale comparison experiments. It was designed with a focus on transparency and reproducibility by putting all the information needed to rerun a comparison in a comprehensive configuration package. Additionally, the wrapper automatically generates most of the statistics and plots contained in this paper.
This section contains a description of the classifier training and testing setup, i.e., the hyperparameter setup as well as the statistics used in the analysis of the results.\bigskip

For setting the training hyperparameters, we adopted the settings as proposed in the original publications on each dataset as cited above. All models were trained by optimizing the categorical cross-entropy loss using the Adam optimizer from \cite{Kingma:2014us}. All training setups including the training duration match the original publications of the respective architectures\cite{Lawhern:2018fb,Lawhern:2016ta,Schirrmeister:2017bv}. \bigskip

To test for significance, a random permutation test was used. This was done by randomly permuting the assignment of test labels to test trials and calculating the resulting accuracy on these randomly permuted labels. By repeating this process $10^6$ times a null distribution was created. Comparing the actual decoding accuracy to the distribution allowed an estimation of how likely it is that a given decoding result was achieved at random, i.e., the significance of the decoding. Following this process, for each recording the significance of the decoding for all four included classifiers was tested. Then, recordings where the p-value exceeded 5\% for all four classifiers were excluded from all further analysis.

As described above the data includes datasets which are difficult to decode. These datasets would add noise to the comparison in cases where all methods yield equally low around-chance-level results. Therefore, only recordings on which at least one of the tested methods achieved a significantly above chance decoding accuracy were further analyzed.

All decoding accuracies mentioned in this paper were calculated as mean class accuracies, by first calculating the accuracy separately for each class and then taking the mean of these accuracies across classes. Thereby the chance level is always at 1/number of classes irrespective of the distribution of examples across classes in the testing data.

The varying difference in task difficulty makes for a broad spread in the distribution of absolute decoding accuracies. To compare these across the four different classifiers, besides the absolute accuracies, we also calculated normalized accuracies by dividing the accuracies obtained with each classifier by the mean across all four different classifiers that we evaluate. Significance of the differences between decoding accuracies was calculated with a two-sided sign test, testing the hypothesis that the difference between to result distributions does not have a zero median.

In addition to comparing the methods' performance as such, we also assessed the correlations of the predictions of the different decoders, both across decoding task examples and on a trial-by-trial basis. One motivation for this analysis was to see whether different methods potentially succeed and fail in different trials; If so, such differences could give hints about the functional properties of different methods and ensembles and combining multiple of the investigated classifiers might allow further decoding accuracy improvements. Therefore, for each method pair, both the prediction dissociation and overlap was computed, i.e., the percentage of trials where both methods either succeed or fail to predict the correct class, as well as the percentage of trials where one method was correct and the other was not.

\begin{figure}[hb]
\centerline{\includegraphics[width=0.5\textwidth]{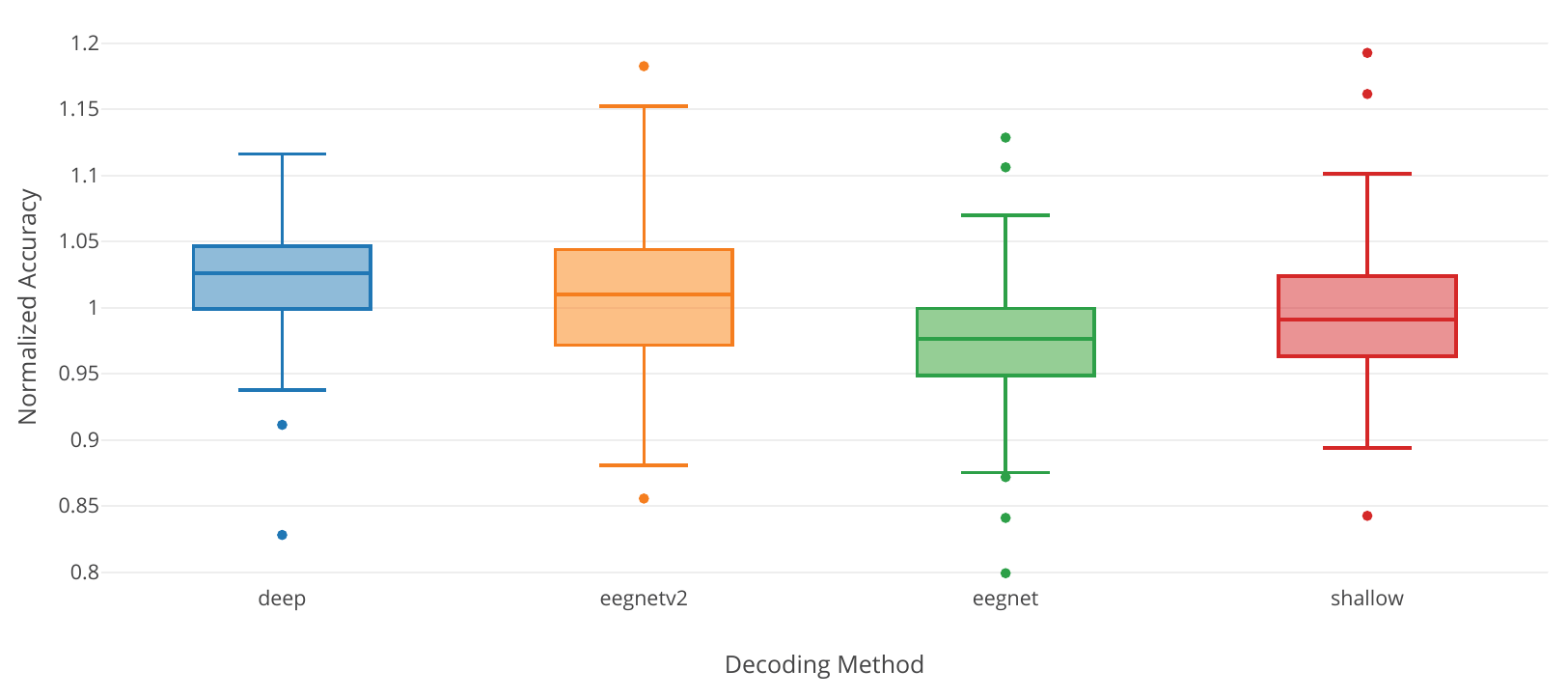}}
\caption{Decoding results normalized to the average performance across classifiers. The Deep4 Network performed better than EEGNet and the Shallow Network, the latter two being almost equal. The new EEGNetv2 architecture improved significantly, now on par with the Deep4 Network.}
\label{figrelative}
\end{figure}

\begin{table}[hb]
\caption{Decoding Results}
\begin{center}
\begin{tabular}{|c|c|c|}
\hline
& \textbf{Mean accuracy}& \textbf{Mean normalized accuracy}\\
\hline
\textbf{Deep Network} & 70.08\% ± 20.92\% & 1.00 ± 0.05\\
\hline
\textbf{EEGNetv2} & 70.00\% ±18.86\% & 1.02 ± 0.08\\
\hline
\textbf{EEGNet} & 67.71\% ± 19.04\% & 0.98 ± 0.06\\
\hline
\textbf{Shallow Network} & 67.71\% ±19.04\% & 0.99 ± 0.06\\
\hline
\end{tabular}
\label{tabdecoding}
\end{center}
\end{table}

\section{Results}
\subsection{Performance comparison}
In 75 out of the 100 decoding task examples, at least one classifier achieved a significantly above chance accuracy (p $<$ 0.05). Only results on these 75 examples are included in the following analyses. Fig. \ref{figrelative} and Table \ref{tabdecoding} give an overview of the overall performance of the four different classifiers compared in the present study. The Deep4 Network performed significantly better than the Shallow Network and first generation EEGNet. The new architecture of EEGNetv2 significantly improved the decoding accuracy compared to its first generation architecture, with a performance of EEGNetv2 not significantly different to that of the Deep4 Network.

\begin{figure*}[ht]
\centerline{\includegraphics[width=\textwidth]{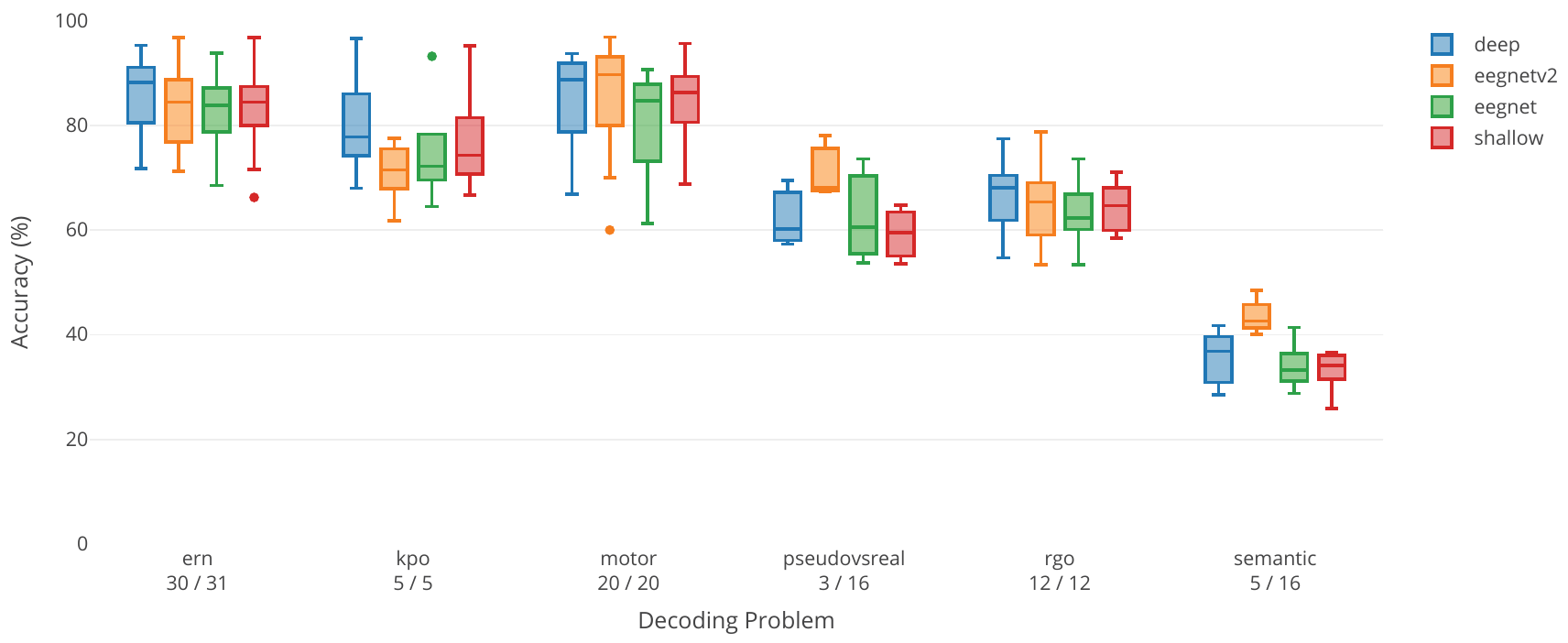}}
\caption{Class mean decoding accuracy per decoding task and decoding method. The overall performance ranking seen in Fig. \ref{figrelative} applied to most individual decoding tasks as well. Note that the box-plots only refer to the examples where at least one of the classifiers yielded a significant result, the number of which is indicated on the bottom together with the total number of examples. For instance in the motor dataset, this included all 20 examples, while in the pseudovsreal semantic decoding problem only three subject had successful decodings with at least one method, illustrating the wide range of difficulty covered by our collection of datasets.}
\label{figdatasets}
\end{figure*}

\begin{figure*}[h!t]
\centering{\includegraphics[width=\textwidth]{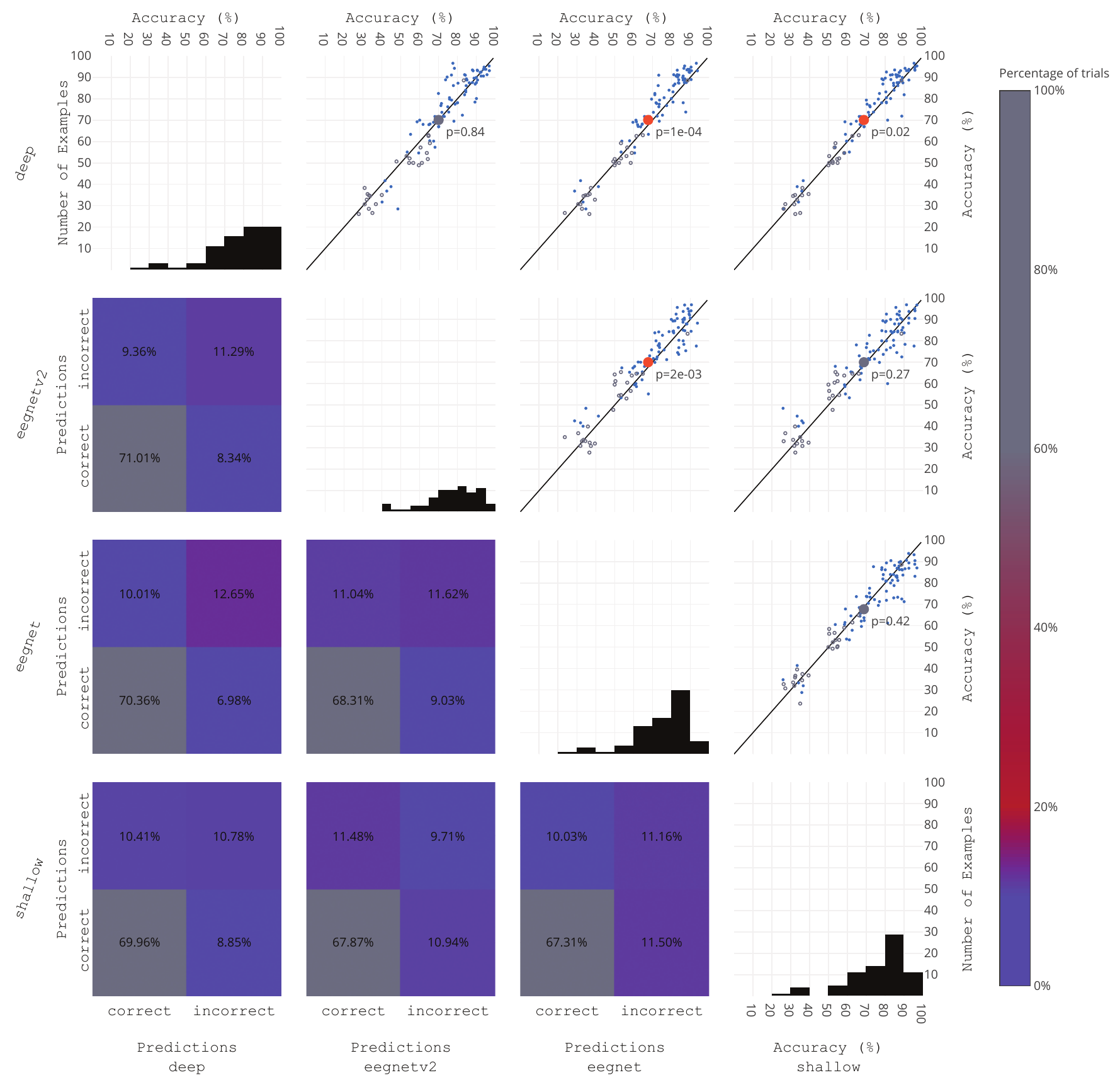}}
\caption{Performance correlations. Rows and columns are decoding methods. The subplots in the diagonal show the histograms of the decoding accuracies of the individual methods. Offdigonal plots show comparisons of method pairs. The upper-right plots are pairwise comparisons of the decoding performance per examples (subjects). Blue dots indicate examples with at least one significant decoding result, open gray circles indicate recordings where no method achieved a significant result. Large dots indicate the mean accuracy over significant examples. The mean was plotted in orange if the overall difference between two methods was also significant (p $<$ 0.05), else in gray, with the corresponding p-value printed next to it. The lower-left plots show color coded matrices of the correlation of correct and incorrect predictions. All X axes of the pairwise plots correspond to the method of the respective column, Y axes to the method of the corresponding row.}
\label{figmatrix}
\end{figure*}

This overall performance ranking is also reflected in the results split according to the six different decoding tasks shown in Fig. \ref{figdatasets}. Fig. \ref{figdatasets} also illustrates the different levels of difficulty of the tasks; in ERN, KPO, motor and RGO, in most of the individual examples significant results were achieved at least by one of the tested methods, in contrast to the much harder semantic tasks. 

As also illustrated by Fig. \ref{figmatrix}, the results of the performance comparison can be summarized as follows: Among the CNNs the Deep4 Network yielded significantly better performance than first generation EEGNet, whereas the Shallow Network’s performance settled in a middle ground between Deep4 Network and EEGNet: The difference between the first and neither of the latter two reached significance. The new architecture of EEGNetv2 significantly improved the decoding accuracy, which is now on par with Deep4 Network's performance.

\subsection{Prediction correlation}
The results of the prediction correlation analysis comparing the four different decoders is shown in Fig. \ref{figmatrix}. Note that the scatter plots also include results on the 25 examples excluded from analysis as gray dots. Fig. \ref{figmatrix} shows that all CNNs disagreed in their predictions on about 15-20\% of the trials. Furthermore, the distribution of correct predictions was relatively balanced. This indicates that ensembles could potentially outperform single CNNs.

\section{Conclusions}
Here we have presented a novel framework for the large-scale evaluation of different deep-learning architectures on different EEG datasets. This framework comprises (i) a collection of EEG datasets, (ii) a collection of different EEG decoding algorithms already published by our lab in an open-source toolbox, and (iii) a wrapper linking the decoders to the data as well as handling structured documentation of all settings and (hyper-) parameters and statistics, designed to ensure transparency and reproducibility. As already done for component (ii), we strive to make the other two components publicly available to the BCI community in the near future as well.   

As an application example of our framework, we have evaluated four CNN architectures. Our comparison also showed that among the CNN architectures evaluated, the Braindecode Deep4 Network and EEGNetv2 performed best. The new EEGNetv2 not only produces similar results compared to our Deep4 Network, but also is architecturally closer to our networks than to the original EEGNet design. On the other hand, the EEGNetv2 architecture is more compact allowing faster training than the Deep4 Network while simultaneously achieving equally high accuracy. We suspect that this could be partially due to the use of separable and depthwise convolutions. We will therefore investigate the possibility of improving the performance of the Braindecode Deep4 Network with these convolution variants. Furthermore, it would also be interesting to investigate potential advantages of deeper networks especially for large EEG datasets. 

Simultaneously, there is an even larger number of other CNN and also RNN architectures that have been proposed for EEG decoding and that are not yet available within our (or any other) framework for large-scale evaluation of such methods. In the future, we will add more deep learning EEG-decoding methods and make them available within our framework. We would hope that methods developed by other researchers in the field would become available in a compatible form as well. We believe that a comprehensive collection of different decoding models that are all compatible with a large collection of EEG test data, as initiated here, could become very helpful for identifying architectures that meet the requirements of specific research and application scenarios.  

In parallel to extending the collection of decoding models, as another future aim we plan to extend our collection of EEG datasets. Possible extensions would be, for example, datasets on the widely used P300 speller paradigms, datasets on "passive BCI" decoding problems such as work load or attention decoding, sleep staging, etc., as well as datasets reflecting a broader range of EEG acquisition techniques and conditions, such as dry EEG or mobile recordings. With a growing data base, a framework as proposed here could also become a useful tool for automatic hyperparameter optimization and architecture search, by providing the large amounts of data typically required by such techniques.

Furthermore, our framework could possibly be developed into a EEG decoding challenge or even an ongoing public benchmark in the likes of the Stanford Question Answering Dataset (SQuAD) \cite{Rajpurkar:2016vf}. Challenges like SQuAD, the ImageNet Competition or the many competitions hosted on the Kaggle platform have benefited progress in many areas where deep learning is applied but are still scarce in the BCI area. In the present study we choose “out of the box” performance evaluation by setting the hyperparameters according to the original publications of the architectures investigated, as an appropriate metric for initial architecture comparison. This is most likely not optimal as the networks may have performed better when using approaches for systematic hyperparameter optimization. Therefore, comparison of such optimized versions of the different proposed architectures would be an interesting future research topic. However, it is unlikely that a single research institution would be able to keep up with all new/updated architecture releases and perform such optimized comparisons. Therefore, a public benchmark would be helpful to alleviate this problem by distributing the workload to all participants while providing comparable results, enabling a comprehensive overview of the state of the art decoding methods' performance. Accordingly, our next step will be to establish such a benchmark utilizing our extensive EEG data collection.

The possibility for evaluation of different deep learning techniques that is large-scale both with respect to the number of network models and datasets included, could be helpful for an informed initial choice of a suitable deep learning architecture also for novel EEG decoding problems. Such an informed initial choice would be especially important when considering deep learning for online BCIs, as in \cite{burget_acting_2017}, where most of the data is acquired after having fixed the decoder architecture. Thus, in summary, the deep learning EEG framework as described here could help to tap the full potential of deep learning for BCI applications.

\section*{Acknowledgment}

We would like to thank the subjects of our EEG experiments for their motivation and commitment. 

\bibliographystyle{IEEEtran}
\bibliography{IEEEabrv,main}

\end{document}